\documentclass[12pt]{article}
\usepackage{epsf,epsfig,cite}

\graphicspath{{figs/}}

\input paperdef

\begin{document}
\begin{titlepage}
\pagestyle{empty}
\baselineskip=21pt
\begin{flushright}
CERN--PH--TH/2007-088\hfill
DCPT/07/52, IPPP/07/26\\
\hfill
UMN--TH--2607/07, FTPI--MINN--07/20 %\\
%arXiv:yymm.nnnn
\end{flushright}
\vskip 0.2in
\begin{center}
{\Large\sc {\bf Light Heavy MSSM Higgs Bosons at Large \boldmath{$\tb$}}}
\end{center}
\begin{center}
\vskip 0.05in
{{\bf J.~Ellis}$^1$, 
{\bf S.~Heinemeyer}$^2$,
{\bf K.A.~Olive}$^{3}$
and {\bf G.~Weiglein}$^{4}$}\\
\vskip 0.05in
{\it
$^1${TH Division, Physics Department, CERN, Geneva, Switzerland}\\
$^2$Instituto de Fisica de Cantabria (CSIC-UC), 
Santander,  Spain \\
$^3${William I.\ Fine Theoretical Physics Institute,\\
University of Minnesota, Minneapolis, MN~55455, USA}\\
$^4${IPPP, University of Durham, Durham DH1~3LE, UK}\\
}
\vskip 0.1in
{\bf Abstract}
\end{center}
\baselineskip=18pt \noindent
%%%%%%%%%%%%%%%%%%%%%%%%%%%%%%%%%%%%%%%%%%%%%%%

{%\small
The region of MSSM Higgs parameter space currently excluded by the CDF
Collaboration, based on an analysis of $\sim 1 \ifb$ of integrated luminosity,
is less than the expected sensitivity.
We analyze the potential implications of the persistence of
this discrepancy within the MSSM, assuming that the soft
supersymmetry-breaking  
contributions to scalar masses are universal, apart from those to the
Higgs masses (the NUHM model).
We find that a light heavy MSSM Higgs signal in the unexcluded part of
the sensitive region could indeed be accommodated in this
simple model, even after taking into account other constraints from
cold dark matter, electroweak precision observables and $B$~physics
observables. In this case the NUHM suggests that
supersymmetric signatures should also be detectable in the near future in
some other measurements such as $\br(B_s \to \mu^+\mu^-)$,
$\br(b \to s \ga)$ and $(g-2)_\mu$,  and $\Mh$ would have to be
very close to the LEP exclusion limit. In addition, the dark matter
candidate associated with this model should be on the verge of detection
in direct detection experiments. 
}

%%%%%%%%%%%%%%%%%%%%%%%%%%%%%%%%%%%%%%%%%%%%%%%%%%%

\vskip 0.15in
\leftline{CERN--PH--TH/2007-088}
\leftline{\today}
\end{titlepage}
\baselineskip=18pt

%%%%%%%%%%%%%%%%%%%%%%%%%%%%%%%%%%%%%%%%%%%%%%%%%%%

%%%%%%%%%%%%%%%%%%%%%%%%%%%%%%%%%%%%%%%%%%%%%%%%%%%%%%%%%%%%%%%%%%%%%%%%%%%%%%%
%%%%%%%%%%%%%%%%%%%%%%%%%%%%%%%%%%%%%%%%%%%%%%%%%%%%%%%%%%%%%%%%%%%%%%%%%%%%%%%

The searches for the bosons appearing in its extended Higgs sector are
among the most promising ways to search for evidence of supersymmetry
(SUSY) at the Tevatron collider. The CDF and D0 Collaborations have already
established important limits on the (heavier) MSSM Higgs bosons,
particularly at large
$\tb$~\cite{CDFbounds,D0bounds,Tevcharged,CDFnew,D0new}.
Recently the CDF Collaboration, investigating the channel
\BE
p \bar p \to \phi \to \tau^+\tau^-, \quad (\phi = h,H,A)~,
\label{pphtautau}
\EE
has been unable to improve these
limits to the extent of the sensitivity expected with the analyzed integrated
luminosity of $\sim 1 \ifb$~\cite{CDFnew}, whereas there is no indication of 
any similar effect in D0 data~\cite{D0new}.%
\footnote{The same analysis shows a deficit of $Z \to \tau^+\tau^-$
  events compared to the Standard Model (SM) expectation. Changing the
  luminosity by one~$\si$ to accomodate this `deficit' 
  would raise the observed rate of $\tau^+\tau^-$ final states around 
  $\MA = 160 \gev$ slightly above the expected rate (by somewhat less than
  one $\sigma$)~\cite{LandsbergTalk}.}% 
~Time will tell whether the CDF effect persists. 
Within the MSSM the channel (\ref{pphtautau}) is enhanced as compared
to the corresponding SM process by roughly a factor of 
$\TQb/((1 + \db)^2 + 9)$~\cite{benchmark3}, where $\tb$ is the ratio of the
two vacuum expectation values, and $\db$ includes loop corrections to
the $\phi b \bar b$ vertex (see \citere{benchmark3} for
details) and is subdominant for the $\tau^+\tau^-$ final state.
Correspondingly, the unexpected weakness of the CDF 
exclusion might be explicable within the MSSM if $\MA \approx 160 \gev$ 
and $\tb \gsim 45$ and, doubtless, also within other theoretical frameworks.

In this paper we investigate whether light heavy Higgs bosons just beyond the
region currently excluded by CDF could be accommodated within
GUT-inspired MSSM scenarios, and what the possible
consequences would be. We consider the constraints imposed by other
measurements, such as the limits on $\br(B_s \to \mu^+\mu^-)$, 
$\br(b \to s \ga)$, $(g-2)_\mu$ and $\Mh$, assuming that
$R$~parity is conserved and that the lightest neutralino $\neu{1}$ 
constitutes the astrophysical dark matter~\cite{WMAP}. 
Whereas we find no solution within the constrained MSSM 
(CMSSM), in which all soft
SUSY-breaking contributions to scalar masses are assumed to
unify at the GUT scale, we find that all the constraints may be
satisfied in the case that universality at the GUT scale 
is relaxed for the scalar Higgs mass parameters (the NUHM
model~\cite{NUHM1,NUHM2,NUHMother}).  
However, we point out that any interpretation of the CDF
effect within the NUHM would be tightly constrained by the other
measurements. Specifically, the constraints are so tight that
one or more of these measurements should display a discrepancy with the
SM, either now or in the near future.

The essence of the argument is as follows. The absence
of exclusion by CDF, 
compared to their expected sensitivity, as mentioned above, 
would correspond to  $\MA (\approx \MH) \sim 160 \gev$,
and a value of $\tb \sim 45$ or greater. Since the
$H/A$ contribution to $\br(B_s \to \mu^+\mu^-) \sim \tan^6\be$,
values of $\tb \gsim 45$ are already excluded for this value of
$\MA$ for substantial portions of the NUHM
parameter space, depending largely on the 
values of $m_{1/2}$ and $m_0$. The parameter space
is so constrained that, in areas which are still allowed,
we expect that a SUSY signal should appear very soon, as we show below. 
We assume $R$-parity conservation, and restrict our attention to the
NUHM with values of the relic CDM 
density $\Omega_{\rm CDM}$ that fall within the range favoured
by WMAP and other astrophysical and cosmological observations.
This restriction imposes important constraints on  
$m_{1/2}$ and $\mu$, the values of $m_0$ and $A_0$ being less
essential. As for $\br(b \to s \ga)$, it is well known that the world
average experimental value currently agrees well with the SM~\cite{hfag}.
In the MSSM,
there are two important contributions with opposite signs, due to
$H^\pm$ and chargino exchanges, respectively. In order for the net MSSM
contribution to be unnoticeable so far, the $H^\pm$ and chargino
exchanges must cancel to a great extent, imposing a relation between the
$H^\pm$ and chargino masses. Since the $H^\pm$  mass is very similar to
the $H, A$ masses, this yields a preferred range of relatively small
values of the chargino mass and hence the soft supersymmetry-breaking
gaugino mass $m_{1/2}$,  favouring in turn a non-negligible
contribution to $(g-2)_\mu$.
The combination of a preferred value for $m_{1/2}$ and the WMAP
constraint then limits the possible range of $\mu$. 
The preference for relatively light sparticles (see also
\citere{ehow4}) translates into a
relatively small value for $\Mh$, and compatibility between the LEP limit
on $\Mh$ \cite{LEPHiggsSM,LEPHiggsMSSM} 
and the upper limit on $\br(B_s \to \mu^+\mu^-)$ \cite{CDFbsmm,D0bsmm}
selects a limited range of $A_0$.  

In preparation for our survey of the NUHM parameter space, we first
recall that in the CMSSM, the electroweak vacuum conditions determine
$|\mu|$ and 
$\MA$ in terms of $\tb$ and the input soft supersymmetry-breaking
parameters. In the NUHM, these are the scalar mass $m_0$ (which is
assumed to be universal, except for the Higgs multiplets)~%
\footnote{We discuss later the implications of relaxing this
  assumption.}%
, the gaugino mass $m_{1/2}$, the trilinear coupling $A_0$ and the degrees of
non-universality of the soft supersymmetry-breaking contributions to the
masses of the two Higgs doublets. In our analysis, we invert the
electroweak vacuum conditions, treating $|\mu|$ and $\MA$ as free
parameters and adjusting the non-universal Higgs mass inputs
accordingly. In the cases of interest this difference in the scalar
masses at the GUT scale is no more than 50\% for low values of $m_0$,
and of order 10\% for higher values of $m_0$. 
The values $\MA = 160 \gev$  and $\tb \ge 45$ are
chosen to match the excess of signal-like events observed in CDF,
and we assume that $\mu > 0$ so as to avoid severe problems with 
$(g-2)_\mu$ and $\br(b \to s \ga)$. Hence our four free parameters are
$m_{1/2}, m_0, A_0$ and $\mu > 0$. However, these are tightly
constrained by other phenomenological limits, as we now discuss.

Since we assume that $R$ parity is conserved, and that the lightest
neutralino $\neu{1}$ constitutes the astrophysical dark
matter~\cite{EHNOS}, we 
impose the requirement that the relic neutralino density falls within the
range allowed by WMAP and other observations: $0.085 < \Omega_{\rm CDM}
h^2 < 0.119$~\cite{WMAP}. As we see later, requiring the relic density
to fall within this narrow range effectively reduces the
dimensionality of the NUHM parameter space. In order to apply the constraints 
on the NUHM parameter space that are provided by $\br(b \to s \ga)$, 
$\br(B_s \to \mu^+\mu^-)$, $(g-2)_\mu$  and $\Mh$, 
we use the following experimental values and theory evaluations (for
more details, specifics on error treatments and an extended list of
references see \citere{ehoww}):
$\br(b \to s \ga)_{\rm exp} = (3.55 \pm 0.24) \times 10^{-4}$~\cite{hfag},
where the theory evaluation is based on \citeres{bsgtheonew,bsgKO2};
$\br(B_s \to \mu^+\mu^-)_{\rm exp} < 10^{-7}$~\cite{CDFbsmm,D0bsmm},
where details about the theory evaluation can be found in~\citere{ourBmumu};
$(g-2)_{\mu,\;{\rm exp-SM}} = (27.5 \pm 8.4) \times 10^{-10}$~\cite{DDDD}
(for a discussion and references to other determinations, see \citere{ehoww}),
and $\Mh > 114.4 \gev$~\cite{LEPHiggsSM,LEPHiggsMSSM}, where the 
theory calculations have been performed with 
{\tt FeynHiggs}~\cite{feynhiggs}.
These observables impose important constraints, as we now 
show~\footnote{Consequences for other observables such as 
$\br(B_u \to \tau \nu_\tau)$ and the $W$-boson mass 
are discussed further below.}.

We consider first the \plane{m_{1/2}}{m_0} shown in panel a) of
\reffi{fig:m0M}, which has $\tb = 45$, $\mu = 370 \gev$ and
$A_0 = -1800 \gev$, as well as $\MA = 160 \gev$.%
\footnote{Here and elsewhere, we assume $m_t = 171.4 \gev$ and 
  $m_b(\mb) = 4.25 \gev$, but our
  results are insensitive to the exact values of $m_t$ and $m_b$, 
  and take into account their current uncertainties, 
  $\de\mt = 1.8 \gev$ and $\de\mb(\mb) = 0.11 \gev$.}%
~The dark (brown) shaded region at low $m_0$ is forbidden, because
there the LSP would be the lighter
stau. The WMAP cold dark matter constraint is satisfied only within the
lighter (turquoise) shaded region. 
To the left of this region, the relic density is too small, due to $s$-channel
annihilation through the Higgs pseudoscalar $A$. As $m_{1/2}$ increases away
from the pole, the relic density increases toward the WMAP range.
However, as $m_{1/2}$ is increased, the neutralino 
acquires a larger Higgsino component and annihilations to pairs of
$W$~and $Z$~bosons become enhanced. To the right of this transition
region, the relic density again 
lies below the WMAP preferred value.  The shaded region here is therefore
an overlap of the funnel and transition regions discussed in~\citere{NUHM2}.

The $\br(B_s \to \mu^+\mu^-)$ constraint is satisfied between the outer 
black dash-dotted lines, labelled $10^{-7}$, representing the
current limit on that branching ratio%
\footnote{
A slightly more stringent upper limit of
$0.93 \times 10^{-7}$ at the $95\%$ C.L.\ has been announced more
recently by the D0~Collaboration~\cite{bsmmD0}. However, applying this
limit would have only a
minor impact on our analysis.
}
.~Also shown are the contours where the
branching ratio is $2 \times 10^{-8}$, close to the sensitivity likely to be
attainable soon by CDF and D0.
Between these two contours, there is a strong cancellation
between  the flavor-violating contributions arising from the Higgs
and chargino couplings at the one-loop level and the Wilson coefficient
counterterms contributing to $\br(B_s \to \mu^+\mu^-)$.

The dash-dotted (red) line shows the contour corresponding to
$\Mh = 114 \gev$, and only the region to the right of this line is
compatible with the constraint imposed by $\Mh$ (it should be kept in
mind that there is still a $\sim 3 \gev$ uncertainty in the
prediction of $\Mh$~\cite{feynhiggs}).
Also shown in pink shading is the
region favoured by $(g-2)_\mu$ at the two-$\sigma$ 
level. The one- and two-$\sigma$ contours for $(g-2)_\mu$
are shown as elliptical dashed and solid black contours, respectively.
The region which is compatible with the WMAP relic density and $\Mh$,
and is also within the two-$\sigma$ $(g-2)_\mu$ experimental bound,
has $\br(B_s \to \mu^+\mu^-) > 2 \times 10^{-8}$. 
The measured value of $\br(b \to s \ga)$ is in agreement with the theory
prediction only to the left
of the solid (green) region. We see, in the space between the $\Mh$
and $b \to s \ga$ exclusions, a slightly diagonal allowed strip of
width $\lsim 200 \gev$ in $m_{1/2}$. 

%%%%%%%%%%%%%%%%%%%%%%%%% F I G U R E %%%%%%%%%%%%%%%%%%%%%%%%%%%%%%%%%%%%%%%%%
\begin{figure}[htb!]
\vspace{20mm}
\begin{center}
\includegraphics[width=.49\textwidth]{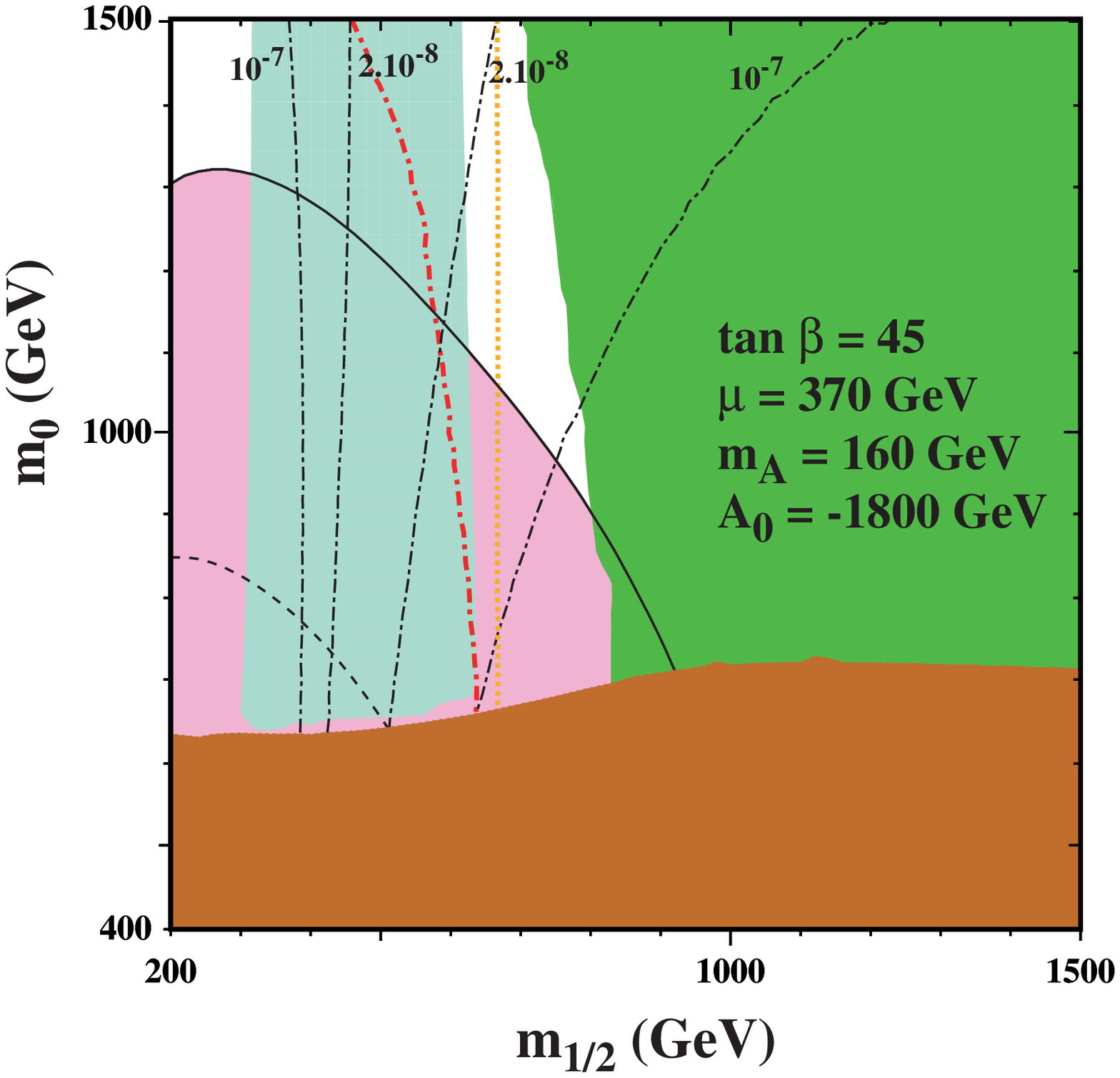}
%\hspace{-25mm}
\includegraphics[width=.49\textwidth]{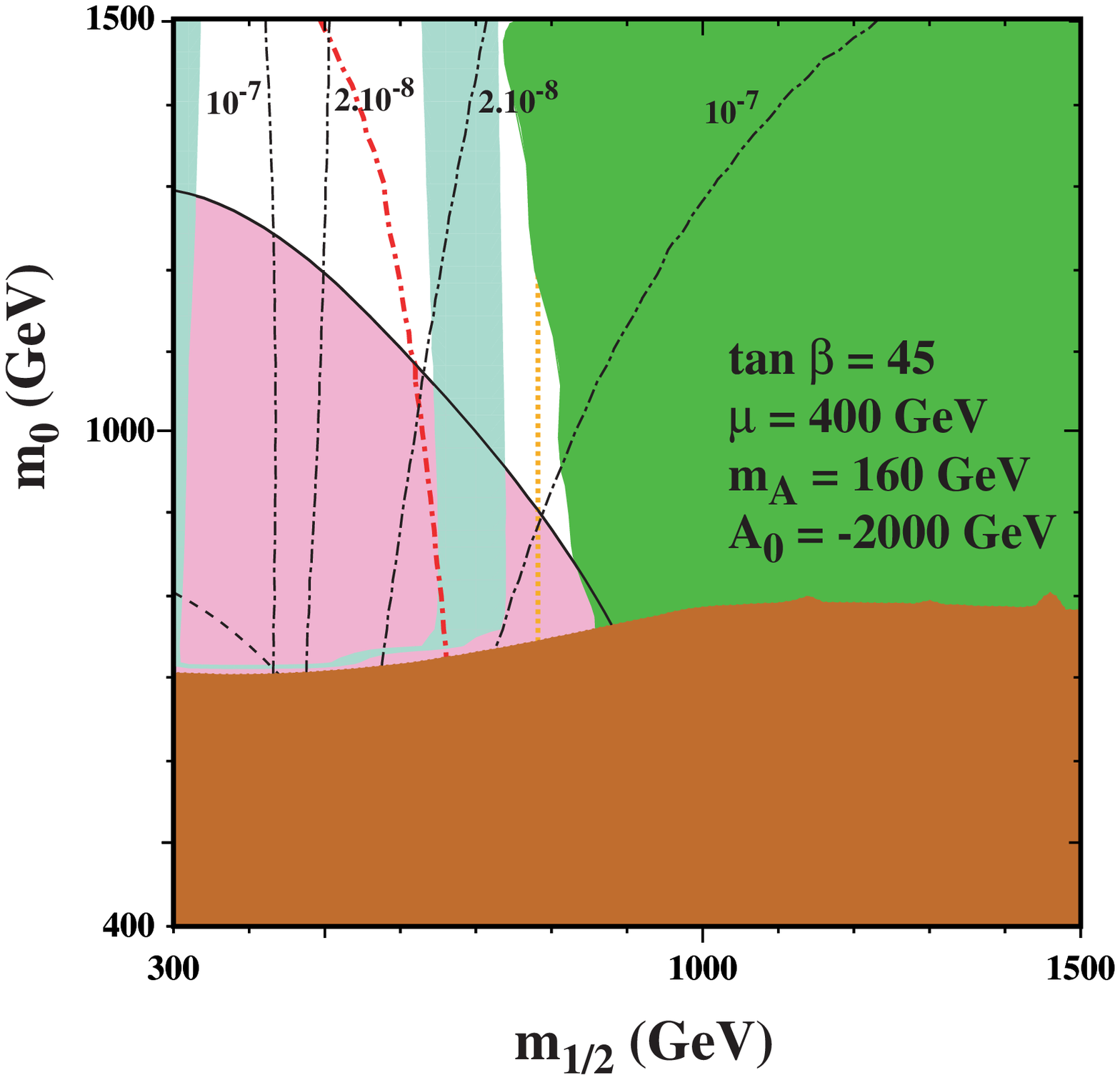}
\includegraphics[width=.49\textwidth]{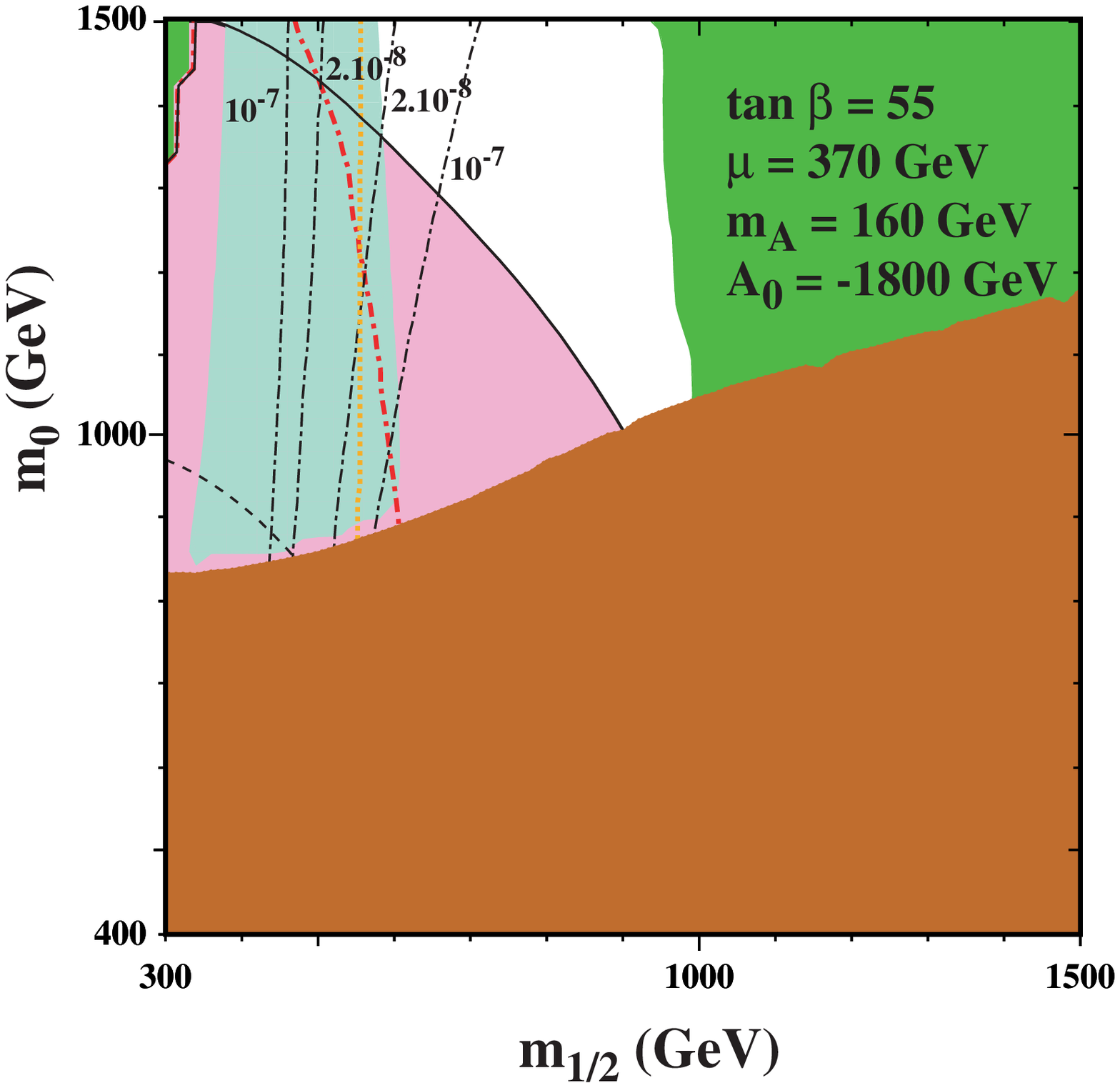}
%\hspace{-25mm}
\includegraphics[width=.49\textwidth]{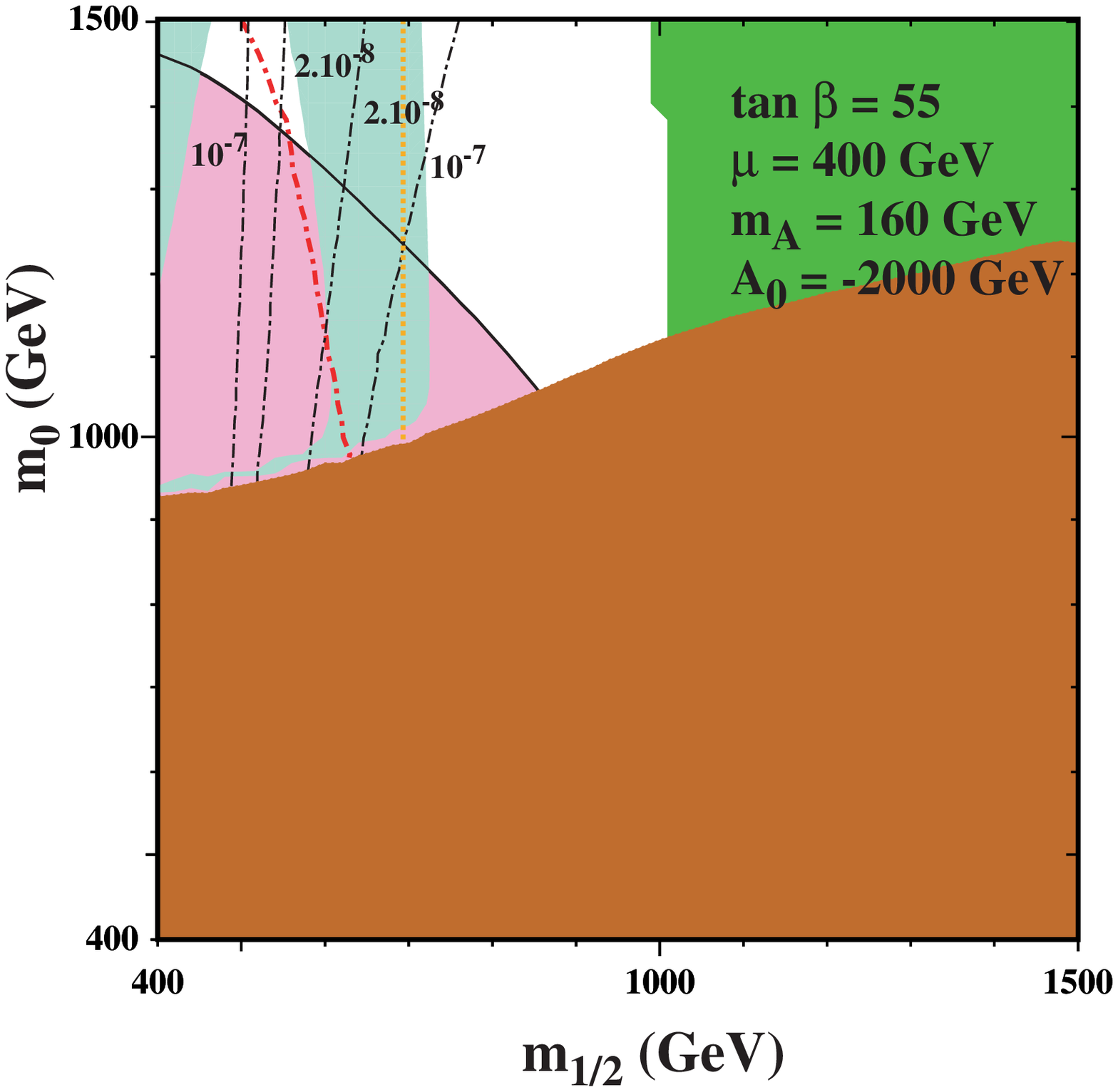}
%\hspace{-5mm}
%\vspace{-25mm}
\caption{%
The NUHM parameter space as a function of $m_{1/2}$ and $m_0$ for 
$\mu = 370 (400) \gev$ and $A_0 = -1800 (-2000) \gev$ in the left
(right) plots. We fix $\MA = 160 \gev$, $\tb = 45 (55)$, $m_t = 171.4 \gev$
and $m_b = 4.25 \gev$ in the upper (lower) plots. 
For the description of the various lines and shaded areas, see the text.
}
\label{fig:m0M}
\end{center}
\vspace{1em}
\end{figure}
%%%%%%%%%%%%%%%%%%%%%%%%% F I G U R E %%%%%%%%%%%%%%%%%%%%%%%%%%%%%%%%%%%%%%%%%

We see that there is a narrow wedge of allowed parameter space in
\reffi{fig:m0M}(a), which has $m_{1/2} \sim 600 \gev$ and 
$m_0 \sim 700$ to 1100~GeV. The $\br(b \to s \ga)$ constraint is satisfied
easily throughout this region, and $(g-2)_\mu$ 
cuts off the top of the wedge, which would otherwise have extended to
$m_0 \gg 1500 \gev$. Within the allowed wedge, $\Mh$ is very close to
the LEP lower limit, and $\br(B_s \to \mu^+\mu^-) > 2 \times 10^{-8}$.
If  $\MA$ were much
smaller ($< 130 \gev$), there would be no wedge consistent 
simultaneously with the $\Omega_{\rm CDM}$,
$\Mh$ and $\br(B_s \to \mu^+\mu^-)$ constraints (the $\Mh$ bound,
however, could then be relaxed due to a weaker $hZZ$ coupling).
At higher values of $\tb$, the allowed region drifts up
to higher values of $m_0$, as is 
shown in  \reffi{fig:m0M}(c) for $\tb = 55$ for the same input values
of $\mu, \MA$, and $A_0$.

We have also considered the potential impact of direct searches for
supersymmetric cold dark matter in the NUHM. 
As has been pointed out previously~\cite{eoss},
these searches are potentially important at small $\MA$ and large
$\tb$, as required in the scenario discussed here (see also~\citere{cdmDD}). 
However, the interpretation of the search limits is less precise than
for the other 
constraints, for two reasons. One is the local density of supersymmetric
cold dark matter, which is usually estimated as $0.3 \gev$/cm$^3$. This
estimate is subject to a systematic uncertainty that is itself uncertain,
but might be $\sim 50 \%$ or so. The second significant uncertainty is in the
hadronic scattering matrix element of the local operator generated by the
short-distance supersymmetric physics. The dominant contribution to
the scattering is spin-independent, and given by the matrix elements in the
nucleons of the scalar quark densities ${\bar q}q$: $q = u, d, s$. These may be
determined from the octet matrix element $\sigma_0$, which is estimated
to be $36 \pm 7 \mev$, and the $\pi$-N scattering matrix element
$\Sigma_{\pi N} = 45 \pm 8 \mev$. These give a range
$y \equiv 2 \langle N | {\bar s} s | N \rangle / 
            \langle N | ({\bar u} u + {\bar d} d) | N \rangle
= 0.2 \pm 0.2$~\cite{oldnp}. Generally speaking, the dark-matter scattering
cross section increases with $y$.

Given the matrix element uncertainties summarized above, we show 
in \reffi{fig:m0M}a) the 1-$\sigma$ lower limit on the calculated value of
the elastic cross section as compared to the CDMS upper limit~\cite{CDMS}.
In the portion of the plane to the left
of the (orange) dotted line, the lower limit on the calculated 
spin-independent elastic cross section is smaller than the CDMS upper bound,
assuming the canonical local density. Whilst we have assumed
$\Sigma_{\pi N} = 45 \mev$, the calculated lower limit effectively
assumes zero strangeness contribution to the proton mass, i.e., $y =
0$. In the region of interest, the lower limit 
on the calculated cross section is about 80\% of the CDMS upper bound, whereas
with a strangeness contribution of $y = 0.2$, the cross section 
would exceed the CDMS bound by a factor of $\sim 3$. 
Thus, if Nature has picked this corner of the NUHM parameter space, we expect 
direct detection of dark matter to be imminent.
We note that the XENON Collaboration has recently announced a
stronger upper limit on the elastic scattering cross
section~\cite{xe10}. Consistency with this limit would further require a
reduction in the local dark matter density (to its lower limit). 
Intriguingly, the
XENON10 experiment has seen some potential signal events that are, however,
interpreted as background. In the following we will show the limits obtained
by CDMS, but the potentially somewhat stronger limits from XENON10 should be
kept in mind.

This example was for the particular values $\mu = 370 \gev$ and 
$A_0 = -1800 \gev$. We now investigate what happens if these values are
varied. 
Panel a) of \reffi{fig:A0mu} explores the \plane{\mu}{A_0} for
$(m_{1/2}, m_0)$ $= (600, 800) \gev$, values close to the lower tip of
the allowed wedge in \reffi{fig:m0M}(a). In this case, the region
allowed by the $\br(B_s \to \mu^+\mu^-)$ constraint is {\it below} the
upper dash-dotted black 
line, and the LEP $\Mh$ constraint is satisfied only {\it above} the
dash-dotted red line. We see that only a restricted range 
$360 \gev < \mu < 390 \gev$ is compatible with the dark matter constraint.
This corresponds to the transition strip where the neutralino is the
appropriate bino/Higgsino combination.  To the left of this strip,
the relic density is too small and to the right, it is too large.
Only a very restricted range of $A_0 \sim - 1600 \gev$ is compatible
simultaneously with the $\Mh$ and $\br(B_s \to \mu^+\mu^-)$ constraints. 
Very large negative values of $A_0$ are excluded as the LSP is the lighter
stau.
On the \plane{\mu}{A_0}, the elastic scattering cross section
is a rapidly decreasing function of $\mu$ and is almost independent of $A_0$.
Indeed, NUHM points excluded by CDMS (or XENON10) generally have low values of 
$\mu$ and $\MA$~\cite{eoss}.
Values of $\mu > 355 \gev$ are compatible with CDMS if the strangeness
contribution to the proton mass is negligible.
 For this choice of parameters,
the entire displayed plane is compatible with $\br(b \to s \ga)$ and
$(g-2)_\mu$. 

%%%%%%%%%%%%%%%%%%%%%%%%% F I G U R E %%%%%%%%%%%%%%%%%%%%%%%%%%%%%%%%%%%%%%%%%
\begin{figure}[t!]
\vspace{20mm}
\begin{center}
\includegraphics[width=.49\textwidth]{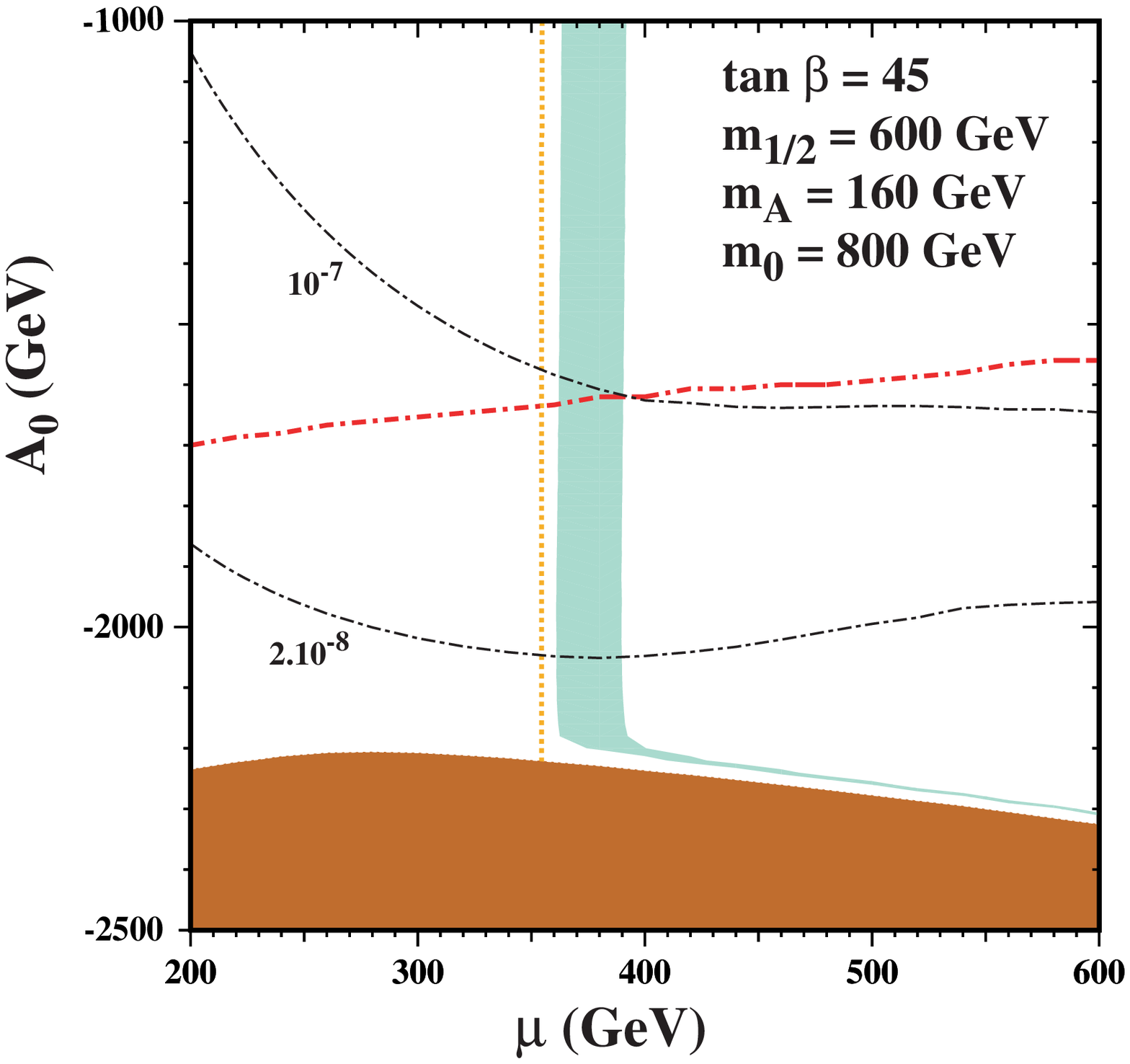}
%\hspace{-25mm}
\includegraphics[width=.49\textwidth]{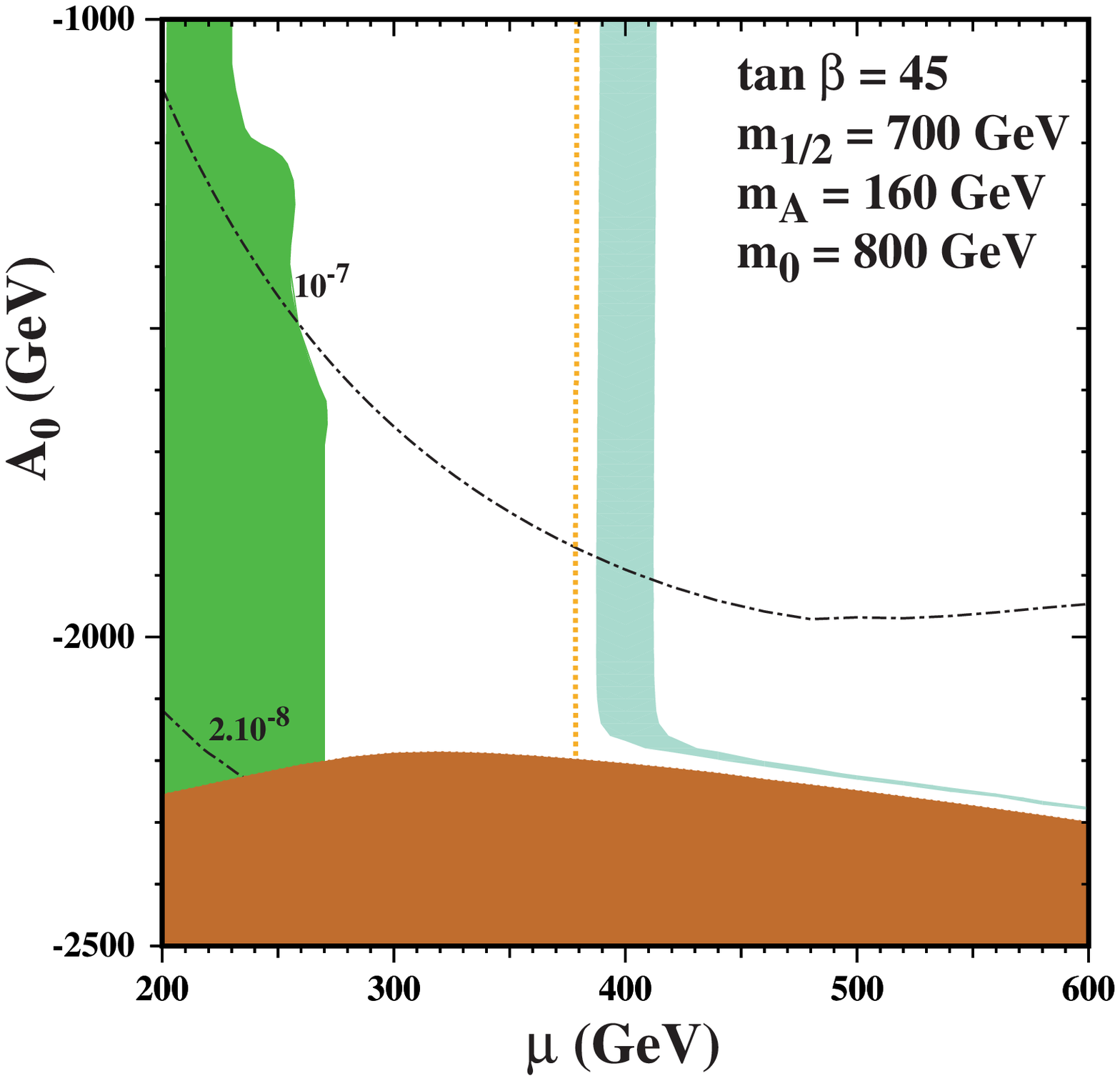}
\includegraphics[width=.49\textwidth]{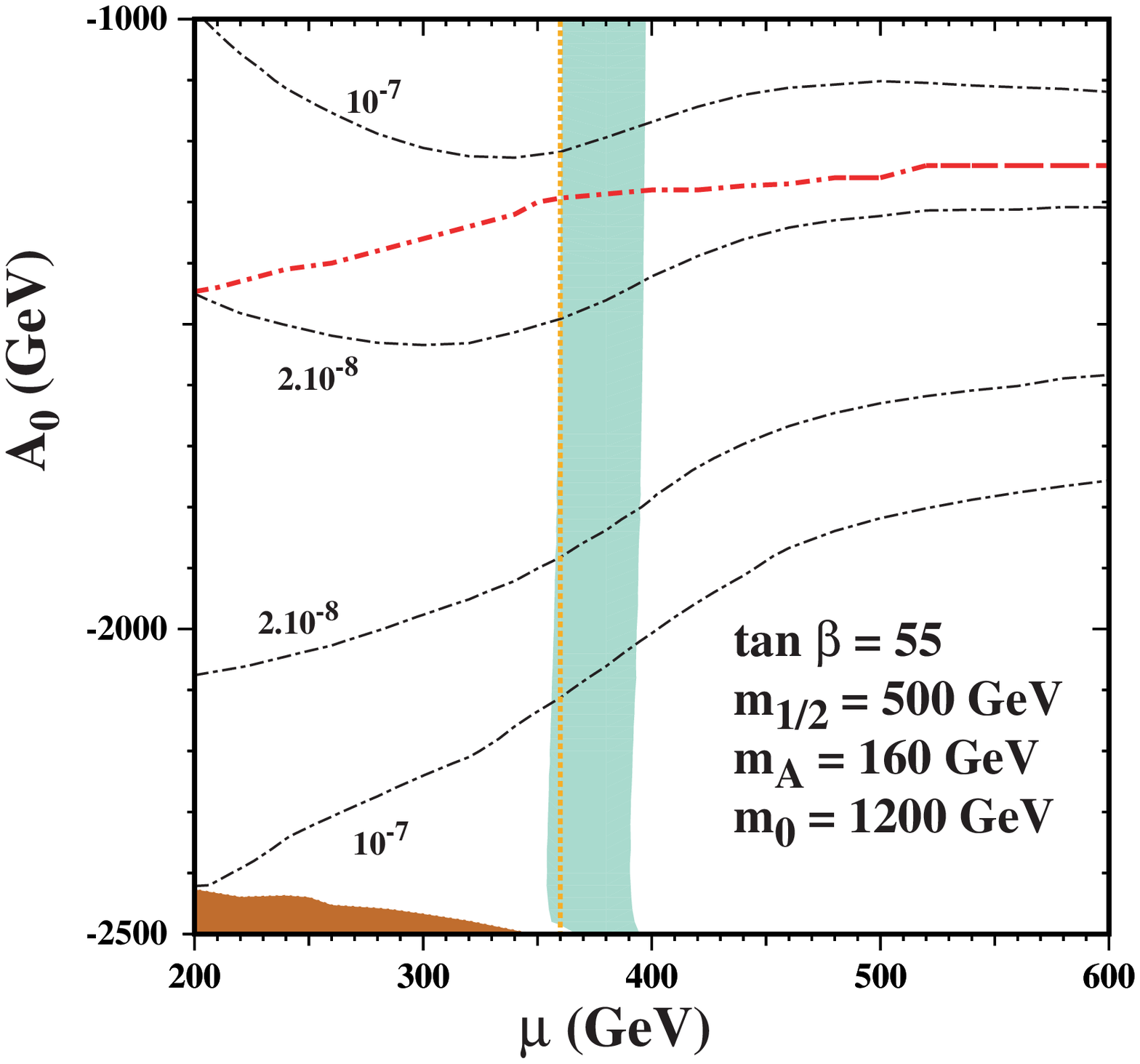}
%\hspace{-25mm}
\includegraphics[width=.49\textwidth]{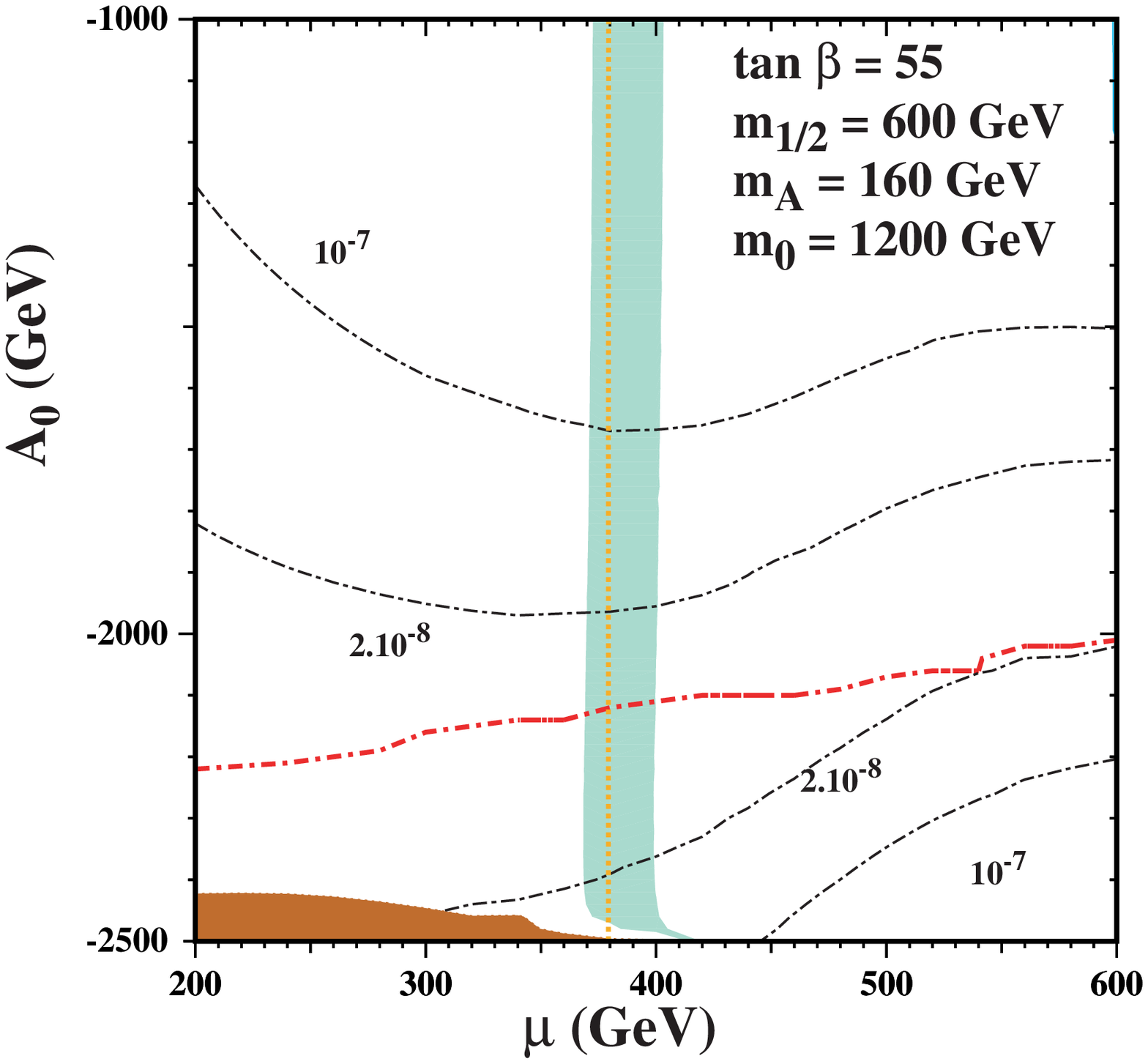}
%\hspace{-5mm}
%\vspace{-25mm}
\caption{%
The NUHM parameter space as a function of $\mu$ and $A_0$ for 
$m_{1/2} = 600 (700) \gev$ and $m_0 = 800 \gev$ in the left
(right) plots. We fix $\MA = 160 \gev$, $\tb = 45 (55)$, $m_t = 171.4 \gev$
and $m_b = 4.25 \gev$ in the upper (lower) plots. 
For the description of the various lines and shaded areas, see the text.
}
\label{fig:A0mu}
\end{center}
\vspace{1em}
\end{figure}
%%%%%%%%%%%%%%%%%%%%%%%%% F I G U R E %%%%%%%%%%%%%%%%%%%%%%%%%%%%%%%%%%%%%%%%%

Panel b) of 
\reffi{fig:A0mu} shows what happens if $m_{1/2}$ is increased to
700~GeV, keeping $m_0$ and the other inputs the same.
Compared to Fig.~\ref{fig:A0mu}(a), we see that the
WMAP strip becomes narrower and shifts to larger $\mu \sim 400 \gev$,  
and that $\br(b \to s \ga)$ starts to
exclude a region visible at smaller $\mu$. If $m_{1/2}$ were to be
increased much further, the dark matter constraint and $\br(b \to s
\ga)$ would no 
longer be compatible for this value of $m_0$. We also see that, by
comparison with \reffi{fig:A0mu}(a), the $\br(B_s \to \mu^+\mu^-)$
constraint has moved to lower $A_0$, but the $\Mh$ constraint has dropped even
further, and $\Mh > 114 \gev$ over the entire visible plane. 
The net result is a region compatible with all the constraints
that extends from $A_0 \sim - 1850 \gev$ down to $A_0 \sim - 2150 \gev$
for $\mu \sim 400 \gev$, with a coannihilation filament extending to
larger $\mu$ when $A_0 \sim - 2200 \gev$. Once again all of the WMAP
strip in this panel is compatible with CDMS.

The larger allowed area of parameter space is reflected in panel b)
of \reffi{fig:m0M}, which has $\mu = 400 \gev$ and $A_0 = - 2000 \gev$,
as well as $\tb = 45$ and $\MA = 160 \gev$ as before. In this
case, we see that a substantial region of the WMAP strip with 
$m_{1/2} \sim 700 \gev$ and a width $\delta m_{1/2} \sim 100 \gev$,
extending from $m_0 \sim 750 \gev$ to higher $m_0$ 
is allowed by all the other constraints. The
$\br(B_s \to \mu^+\mu^-)$ and $\Mh$ constraints have now moved to
relatively low values of $m_{1/2}$, but we still find 
$\br(B_s \to \mu^+\mu^-) > 2 \times 10^{-8}$ and $\Mh$ close to the
LEP lower limit. We note that the $(g-2)_\mu$ and 
$\br(b \to s \ga)$ constraints now disfavour a bigger fraction of the
parameter space with large $m_{1/2}$, but their only effect here 
is that $(g-2)_\mu$
truncates the allowed region at $m_0 \sim 1050 \gev$.  Once again,
the allowed region is compatible with CDMS (to the left of the (orange)
dotted line) provided the strangeness contribution to the proton mass is
small.  

Panels c) and d) of  \reffi{fig:m0M} show the constraints on the
\plane{m_{1/2}}{m_0} for $\tb = 55$ for the same set of input
parameters used in panels a) and b). 
The dominant effects at higher $\tb$ are the shift to the left 
of the contours of $\br(B_s \to \mu^+\mu^-)$ and the exclusion of a much
larger portion of the plane at low $m_0$ due to a stau LSP.
Because of the strong dependence on $\tb$, the 
$\br(B_s \to \mu^+\mu^-)$ constraint essentially excludes all values of
$m_{1/2} \gsim 700 \gev$ in panel c) and $\gsim 750 \gev$ in panel
d). Nevertheless, a viable portion of parameter space remains in tact at
lower $m_{1/2}$ and higher $m_0$. At higher $\tb$, the CDMS bound
also becomes stronger and restricts $m_{1/2} \lsim 560 \gev$ in panel c)
and $\lsim 700 \gev$ in panel d).  
Thus in panel c), we are forced into a small triangular region
centered at $m_{1/2} \sim 530 \gev$ and $m_0 \sim 1350 \gev$ bounded by
CDMS, $\Mh$, and $(g-2)_\mu$. There, the branching ratio, 
$\br(B_s \to \mu^+\mu^-)$ is near its minimum value of 
$2 \times 10^{-9}$ due to the strong cancellation discussed above. 
In panel d) we are left with a significantly larger quadrilateral region. 

The analogous \plane{\mu}{A_0}s are shown in \reffi{fig:A0mu}c) and d).
In panel c), only a small portion of the WMAP strip around 
$A_0 \sim -1200 \gev$ is compatible with both $\br(B_s \to \mu^+\mu^-)$
and $\Mh$, while the range in panel d) is considerably larger.  In both
cases, the CDMS limit is stronger and begins to cut into the WMAP strip.

We have surveyed systematically the allowed region of parameters in the
\plane{m_{1/2}}{m_0} for $\tb$  between 45 and 55, varying the inputs
$\mu$ and $A_0$. 
To summarize the regions in the NUHM parameter space which are compatible
with CDM and all phenomenological constraints {\em and} a light heavy
Higgs, we first 
recall that the values of $m_{1/2}$ compatible with WMAP depend on
$\mu$, but depend less on $m_0$. Specifically, as $\mu$ increases, the
preferred region of the \plane{m_{1/2}}{m_0} is a near-vertical strip that
moves to larger $m_{1/2}$, which is truncated at low $m_0$ just above the 
region where the lighter stau is the LSP. For $\mu < 350 \gev$, there is no
WMAP-compliant region compatible with the LEP lower limit on $\Mh$, but a
small allowed region appears for $\mu$ slightly below 
$370 \gev$. Typical values for 
$m_{1/2}$ are $500 - 600 \gev$ at this value of $\mu$ and $A_0$ must be
below $-1300 \gev$. When $\mu = 400 \gev$, the WMAP-compatible
strip moves to larger $m_{1/2}$, and there are allowed 
regions of the \plane{m_{1/2}}{m_0} for $A_0 \sim -1600 \gev$ to 
$\sim - 2400 \gev$. As $\mu$ increases further, the WMAP-compatible strip moves
to even larger $m_{1/2}$, and the only allowed region is a small piece
of coannihilation strip close to the boundary with the stau LSP
region. This region is negligible for $\mu > 500 \gev$. Combining all the
allowed values of $\mu$ and $A_0$, we find that only a small portion of
the \plane{m_{1/2}}{m_0} can ever be compatible with all the
constraints. It is roughly triangular in shape, with vertices 
$(m_{1/2}, m_0) = (500, 1400),  (700, 700)$ and $(800, 900) \gev$.

We now analyse the dependence of the results on the assumption of scalar-mass
universality. The assumptions that all squarks with the same electroweak
quantum numbers have identical input soft supersymmetry-breaking scalar
masses is motivated in general by the suppression of flavour-changing
neutral interactions. Specifically, the upper limit on 
$\br(B_s \to \mu^+\mu^-)$ and the agreement of 
$\br(b \to s \ga)$ with the SM would require a certain degree of
fine-tuning in the presence of large squark 
mass non-universality. On the other hand, the relative locations of both
the WMAP strip and the $(g-2)_\mu$ constraint depend on the relationship
between the soft SUSY-breaking squark and slepton scalar
masses. If the slepton masses are {\it decreased} relative to the 
squark masses (which we continue to denote by $m_0$), the lower ends of
the WMAP strips in \reffi{fig:m0M} will {\it rise} (due to relatively
ligher staus), as will the
$(g-2)_\mu$ contours, {\it raising} the preferred range of $m_0$.
Conversely, if the slepton masses are {\it increased} relative  to the
squark masses, the preferred range of $m_0$ will be {\it lower}. 
Thus, neglecting the universality between squark and slepton masses
in general enlarges the allowed region of parameter space that is
compatible with the experimental constraints.

We now discuss the possible phenomenological signatures of a scenario
with $\MA = 160 \gev$ and $\tb \gsim 45$ within the NUHM.
The interplay of the various constraints in \reffi{fig:m0M} implies:
\begin{itemize}
\item[(i)]
The predicted value of $\br(B_s \to \mu^+\mu^-)$ in the allowed region
is generally $ > 2 \times 10^{-8}$. Thus, this channel may offer good
prospects within the near future for either supporting or contradicting
the NUHM interpretation of the weaker CDF $\tb$ bound, as compared to the
expected sensitivity. 
\item[(ii)]
We find that $\Mh$ must be very close to the LEP lower limit, i.e.\ in
the range 
where LEP observed a couple of Higgs-like events~\cite{LEPHiggsMSSM}.
This part of parameter space could be probed at the Tevatron with 
$8 \ifb$ of integrated luminosity. Due to the large value of $\tb$ and the
small mass of the $A$~boson, the rates of $h$ decays into bottom quarks
and tau leptons 
are enhanced as compared to the SM.
\item[(iii)]
The predicted value of $\br(b \to s \ga)$ in the allowed region is 
$\sim 4.6 \times 10^{-4}$, which is about one~$\si$ above the current
experimental value (if the experimental and theory errors are added
linearly). Consequently, an improvement in the present
theoretical uncertainty might enable a discrepancy to appear between 
$\br(b \to s \ga)$ and the SM value.
\item[(iv)]
The discrepancy between the experimental measurement of $(g-2)_\mu$ and
the  SM calculation can easily be explained in this scenario, 
although the
$(g-2)_\mu$ discrepancy should be somewhat smaller than the
current central value. However, a much smaller discrepancy
(corresponding to larger $m_0$ values) could also be accomodated.
\item[(v)]
Confronting the prediction for $\br(B_u \to \tau \nu_\tau)$~\cite{BPOtheo}
with the measurement from Belle and B{\sc A}B{\sc AR}~\cite{BTNexp} 
already yields interesting constraints on the charged Higgs-boson mass
as a function of $\tb$, although the present experimental errors are
still very large, see e.g.\ \citere{BTNSUSYint}. In the scenario
considered here $\br(B_u \to \tau \nu_\tau)$
is predicted to be
relatively low as compared to its SM value, where the ratio of MSSM/SM is
$\sim 0.33$. This is about one~$\si$ below the current
central value.
\item[(vi)]
The ratio of the $B_s$ mass difference to the SM prediction is
close to unity, $\sim 0.91$~\cite{BPOtheo}. In view of the theoretical
uncertainties, it 
will be difficult to establish such a small deviation.
\item[(vii)]
The $W$~boson mass is predicted to be~\cite{MWweber,PomssmRep}
relatively low, $\MW \sim 80.367 \gev$, i.e.\ again about one~$\si$
below the current central experimental value.
\item[(viii)]
The masses of the other SUSY particles are estimated to be
$\mneu{1} \sim 210$--$270 \gev$,
$\mstaue \sim 310$--$800 \gev$,
$\mcha{1},  \mneu{2} \sim 340$--$390 \gev$,
$\mneu{3} \sim 370$--$410 \gev$,
$\mcha{2}, \mneu{4} \sim 480$--$590 \gev$,
$\mstauz, m_{\tilde{\nu}_\tau} \sim 770$--$1120 \gev$,
$m_{\tilde e_R} \sim 870$--$1250 \gev$,
$m_{\tilde e_L}, m_{\tilde{\nu}_e} \sim 960$--$1310 \gev$,
$\mgl \sim 1270$--$1570 \gev$,
and for the squarks $\mste \sim 1250 \gev$, 
$\mstz, \msbe \sim 1450 \gev$,
$\msbz \sim 1550 \gev$, 
$m_{\tilde u_{L,R}},  m_{\tilde d_{L,R}} \sim 1600 \gev$,
each with uncertainties $\sim 10 \%$.
The sparticle spectrum is not particularly light, and at the Tevatron no
further SUSY particle discoveries could be expected. On the other hand, 
the strongly-interacting sparticles should mostly be within
reach of the LHC, and many weakly-interacting sparticles should
be visible in their cascade decays.
At the ILC, depending on the center-of-mass energy, the ligher
neutralinos, charginos and staus could be produced.
\item[(ix)] Direct detection experiments should see a signal at
current sensitivities. The lack of a signal in the CDMS experiment could
be explained by a reduced strangeness contribution to the proton mass.  
The reported XENON10 limit would further require a reduction in the local
halo density (to its lower limit).

\end{itemize}

Performing a $\chi^2$ fit for nine precision and $B$-physics observables
along the lines of \citere{ehoww} yields a total 
value of $\chi^2_{\rm tot} \sim 9-10$ in the allowed part of the NUHM
parameter space, where even
slightly smaller values can be found for $\Mh \lsim 114 \gev$.%
\footnote{It should be kept in mind that the $\Mh = 114 \gev$ lines 
shown in the plots are roughly the LEP 95\% C.L. exclusion limits, 
not the 2\,$\si$ bound. In applying this bound the current theory
uncertainty in the $\Mh$ prediction of $\sim 3 \gev$ needs to be taken
into account.}%

Very likely the weaker CDF $\tb$ bound from the search for heavy Higgs
bosons compared to its
expected sensitivity is due to a statistical fluctuation that will eventually
evaporate. Nevertheless, it is interesting to know whether a signal at this
level could be accommodated within the MSSM. We have shown that
$\MA \sim 160 \gev$ is possible for $\tb \gsim 45$
within the NUHM, though it stretches
various experimental constraints to their limits. Correspondingly, small
improvements in some of these measurements, e.g., of 
$\br(B_s \to \mu^+\mu^-)$ or of the dark matter scattering rate, 
would either exclude such a signal in
the NUHM framework, or else provide supporting evidence. One way or
another, it should be possible soon to cast light on the interpretation
of the CDF search.

\subsection*{Acknowledgements}
We would like to thank Patricia Ball, Prisca Cushman, and Athanasios Dedes 
for useful conversations. 
The work of K.A.O.\ was partially supported by DOE grant DE-FG02-94ER-40823. 
The work of S.H.\ was partially supported by CICYT (grant FPA2006--02315).
Work supported in part by the European Community's Marie-Curie Research
Training Network under contract MRTN-CT-2006-035505
`Tools and Precision Calculations for Physics Discoveries at Colliders'.

%%%%%%%%%%%%%%%%%%%%%%%%%%%%%%%%%%%%%%%%%%%%%%%%%%%%%%%%%%%%%%%%%%%%%%%%%%%%%%%
%%%%%%%%%%%%%%%%%%%%%%%%%%%%%%%%%%%%%%%%%%%%%%%%%%%%%%%%%%%%%%%%%%%%%%%%%%%%%%%

\end{document}

%%%%%%%%%%%%%%%%%%%%%%%%%%%%%%%%%%%%%%%%%%%%%%%%%%%%%%%%%%%%%%%%%%%%%%%%%%%%%%%
%%%%%%%%%%%%%%%%%%%%%%%%%%%%%%%%%%%%%%%%%%%%%%%%%%%%%%%%%%%%%%%%%%%%%%%%%%%%%%%